\documentclass[aip, reprint, nofootinbib]{revtex4-1}
\usepackage{graphicx}
\usepackage{glossaries}
\usepackage{amsmath}
\usepackage{amssymb,amsmath}
\usepackage[usenames,dvipsnames]{color}
\usepackage[margin=.57in]{geometry}
\usepackage{braket}
\usepackage{enumitem}
\usepackage{mathtools}

\usepackage[english]{babel} 

\usepackage{ntheorem}
\theoremseparator{:}
\usepackage{ulem}

\usepackage{filecontents}
\pdfoutput=1

\begin{document}
\normalem

\title{Quantized Hall Effect Phenomena and Topological-Order in 4D Josephson Junction Arrays in the Vicinity of a Quantum Phase Transition}

\author{Caroline S. Gorham}
\email{caroling@cmu.edu}
\affiliation{Department of Materials Science and Engineering, Carnegie Mellon University, Pittsburgh, PA 15213, USA}

\author{David E. Laughlin}
\email{laughlin@cmu.edu}
\affiliation{Department of Materials Science and Engineering, Carnegie Mellon University, Pittsburgh, PA 15213, USA}

\begin{abstract}
Recently, generalizations of quantum Hall effects (QHE) have been made from 2D to 4D and 8D by considering their mathematical frameworks within complex ($\mathbb{C}$), quaternion ($\mathbb{H}$) and octonion ($\mathbb{O}$) compact (gauge) Lie algebra domains. Just as QHE in two-dimensional electron gases can be understood in terms of Chern number topological invariants that belong to the first Chern class, QHE in 4D and 8D can be understood in terms of Chern number topological invariants that belong to the $2^\text{nd}$ and $4^\text{th}$ Chern classes. It has been shown that 2D QHE phenomena are related to topologically-ordered ground states of Josephson junction arrays (JJAs), which map onto an Abelian gauge theory with a {periodic} topological term that describes charge-vortex coupling. In these 2D JJAs, magnetic point defects and Cooper pair electric charges are dual to one another via electric-magnetic duality (Montonen-Olive). This leads to a quantum phase transition between phase-coherent superconductor and dual phase-incoherent superinsulator ground states, at a ``self-dual'' critical point. In this article, a framework for topological-ordering of Bose-Einstein condensates is extended to consider four-dimensional quaternion ordered systems that are related to 4D QHE. This is accomplished with the incorporation of a non-Abelian topological term that describes coupling between third homotopy group point defects (as generalized magnetic vortices) and Cooper pair-like charges. Point defects belonging to the third homotopy group are dual to charge excitations, and this leads to the manifestation of a quantum phase transition between orientationally-ordered and orientationally-disordered ground states at a ``self-dual'' critical point. The frustrated ground state in the vicinity of this ``self-dual'' critical point, are characterized by global topological invariants belonging to the $2^{nd}$ Chern class. 
\end{abstract}
\maketitle

\section{Introduction}

Different states of matter have different physical properties based on their type and degree of order. In addition to conventional Landau ordered states of matter, obtained by spontaneous symmetry breaking and characterized by local order parameters, topologically-ordered states of matter exist that are described by global topological invariants~\cite{lohse_exploring_2018} and that are therefore {robust against perturbations}. Striking examples of {topologically ordered phases} are quantum Hall effect (QHE) systems, for which the most notable example occurs in two-dimensional electron gases (2DEGs) at very low temperatures and in a magnetic field. In these 2D QHE systems, precise quantization of Hall conductance has topological origins that can be understood in terms of topological invariants known as the Chern numbers -- that belong to the $1^\text{st}$ Chern class. Notably, 2D QHE phenomena also play a major role in understanding the topologically-ordered frustrated  ground states of Josephson junction arrays~\cite{diamantini_gauge_1996, diamantini_topological_2008} (JJAs). 

\begin{figure*}
  \centering
\includegraphics[width=\textwidth]{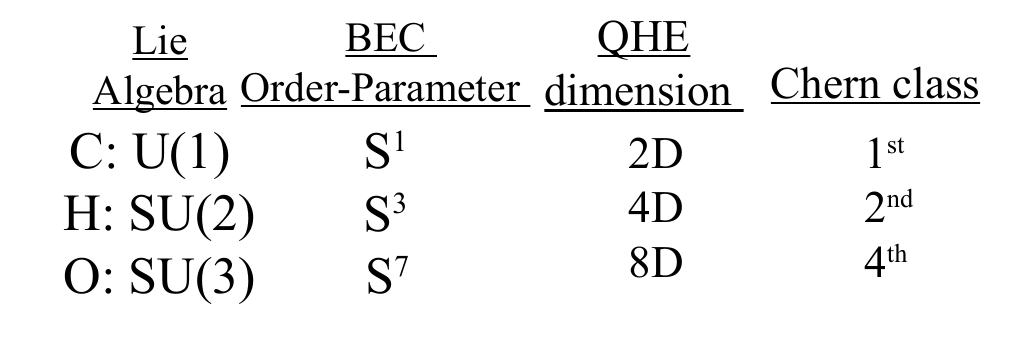}
\caption{Complex numbers ($\mathbb{C}\cong S^1\in \mathbb{R}^2$) describe the $U(1)$ group, and $SU(n+1)$ groups are described by quaternion ($n=1$, $\mathbb{H}\cong S^3\in \mathbb{R}^4$) and octonion ($n=2$, $\mathbb{O}\cong S^7\in \mathbb{R}^8$) numbers. These division algebras are the fundamental structures for QHE~\cite{bernevig_eight-dimensional_2003} in 2D, 4D and 8D -- for which precise quantization of Hall conductance is related to the $m^{th}$ Chern class.}
\label{fig:chern}
\end{figure*}

Generalizations of QHE from 2D to 4D and 8D have been made by considering the fundamental structures of QHE as given by the division Lie algebras~\cite{zhang_four-dimensional_2001, bernevig_eight-dimensional_2003}: complex ($\mathbb{C}$), quaternion ($\mathbb{H}$) and octonion ($\mathbb{O}$). These division Lie algebras have the symmetry of compact odd-dimensional spheres: $\mathbb{C}: U(1)\cong S^1$, $\mathbb{H}: SU(2)\cong S^3$ and $\mathbb{O}: SU(3)\cong S^7$ which characterize the degree of Bose-Einstein condensation in particle systems (with relevant symmetry) below a critical temperature (e.g., superfluidity Bose-Einstein condensates are characterized by a complex order parameter). Alternatively, the higher-dimensional division algebras with the group structure $SU(n+1)$ can be represented by $m-$dimensional ($m=2n$) complex projective spaces $\mathbb{C}^m$:  $SU(2):\mathbb{H}\cong \mathbb{C}^2$ and $SU(3):\mathbb{O}\cong \mathbb{C}^4$. Just as QHE in 2DEGs are understood using $1^\text{st}$ Chern class topological invariants, QHE in systems of particles with $SU(n+1)$ symmetry occur in $2m$-dimensions and are understood by Chern topological invariants that belong to the $m^{th}$ Chern class. Figure~\ref{fig:chern} summarizes these statements.

     \begin{figure}[b!]
  \centering
\includegraphics[scale=.7]{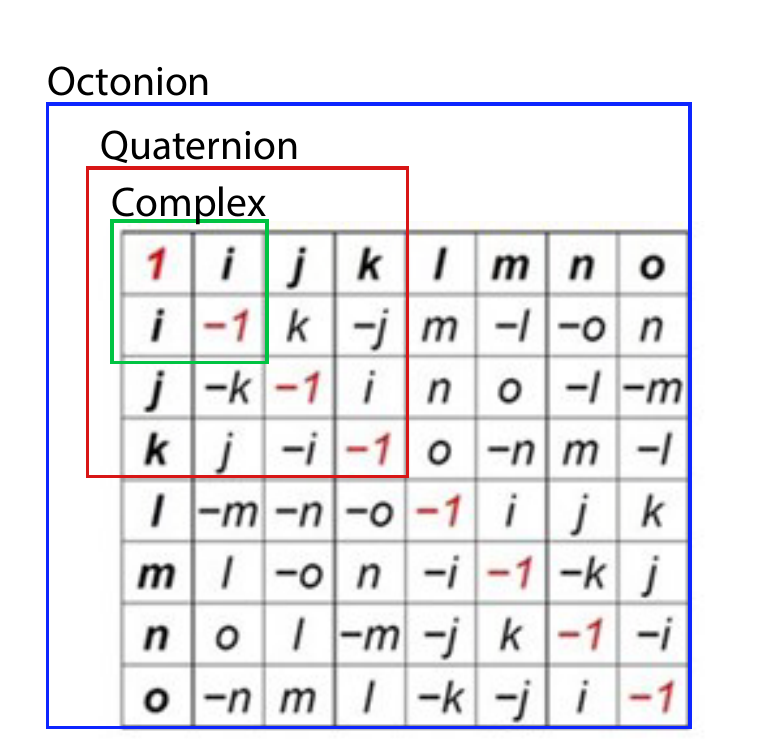}
\caption{Octonion ($\mathbb{O}$) group multiplication table, with quaternion ($\mathbb{H}$) and complex ($\mathbb{C}$) group multiplication tables as subgroups. The $\mathbb{C}$ group is Abelian, and the $\mathbb{H}$ and $\mathbb{O}$ groups are non-Abelian (i.e., group operation is non-commutative).  }
\label{fig:oct_mult}
\end{figure}

In this article, in particular, 4D QHE phenomena in systems of particles with $SU(2)$ symmetry are discussed.  Such higher-dimensional 4D QHE systems, associated with the $2^\text{nd}$ Chern class, have been considered theoretically~\cite{zhang_four-dimensional_2001, bernevig_eight-dimensional_2003, kraus_four-dimensional_2013} and have recently been observed in thin-film superlattices that act as two interacting 2D QHE topological charge pumps (Ref.~\onlinecite{lohse_exploring_2018}) such that the 4D QHE may be described using two coupled $T^2$ surfaces (instead of a single $T^4$ surface). This convenient representation of the 4D QHE is related to the fact that four-dimensional quaternion numbers may be modeled as pairs of complex numbers\footnote{Construction of a quaternion as a pair of complex numbers is a generalization of the construction of a complex number as a pair of real numbers in the complex plane:
\begin{equation}
\mathbb{C} :\{ \textbf{z} = r_0 + \hat{i}r_1  \,\,\, | \,\,\, r_0^2 + r_1^2 = R^2 \} \,\, 
\end{equation}
where $R$ is the radius of a circle ($S^1$), $r_0$, $r_1$ are real numbers and $\hat{i}$ is a pure imaginary complex number (i.e., $\hat{i}^2=-1$).}, as $\mathbb{H}=\mathbb{C}^2$. Letting $\mathbb{C}^2$ be a two-dimensional vector space over the complex numbers, a quaternion number takes the form:
\begin{equation}
\mathbb{H} : \{ \textbf{q}=(\textbf{z}_1, \textbf{z}_2)\in \mathbb{C}^2 \,\,\, | \,\,\, |z_1|^2 +  |z_2|^2 = R^2 \},
\end{equation}
where $\textbf{z}_1 \equiv x_1 + \hat{i} p_1$ and $\textbf{z}_2 \equiv x_2 + \hat{i} p_2$ and $\hat{i}$ is a pure imaginary complex number. Any quaternion vector in $\mathbb{C}^2$ may be constructed in terms of basis elements $1$ and $\hat{j}$: 
\begin{equation}  
\textbf{q} = (x_1 + p_1\hat{i})1 + (x_2+p_2\hat{i})\hat{j}.
\label{eqn:quaternion_number}
\end{equation} 
Any two vectors can be multiplied using a distributive law by defining $\hat{j}^2 = -1$ and $\hat{i}\hat{j}=-\hat{j}\hat{i}$, and defining the product $\hat{i}\hat{j}=\hat{k}$ leads to quaternion multiplication rules (Figure~\ref{fig:oct_mult}).

Just as numbers that belong to the complex group may be written using Euler's formula: $\textbf{z} = |z|e^{\hat{i}\theta}$, where $\theta\in[0,2\pi]$ is the single scalar phase angle that parameterizes $S^1\in \mathbb{R}^2$, quaternion numbers (Eqn.~\ref{eqn:quaternion_number}) may be described using an extension of Euler's formula:  $\textbf{q}=|q|e^{\hat{\tau}\theta}$, where $\hat{\tau}  = \cos\theta_1\hat{i} + (\sin\theta_1\cos\theta_2)\hat{j} + (\sin\theta_1\sin\theta_2) \hat{k}$ is a pure imaginary quaternion (i.e., $\hat{\tau}^2=-1$), and ($\theta$, $\theta_1)\in ([0,\pi], [0,\pi])$ and $\theta_2\in[0,2\pi]$ are the three scalar phase angles that parameterize $S^3\in\mathbb{R}^4$. 

These parameterizations point towards the only kind of topological defects that can arise in type-II charged Bose-Einstein condensates that are characterized by complex (i.e., superconductors) or quaternion orientational order parameters~\cite{toulouse_principles_1976, gorham_topological_2019}. Specifically, these topological defects are singularities in the order parameter field that belong to the fundamental (first) and third homotopy groups respectively (i.e., $\pi_1(S^1)$ and $\pi_3(S^3)$) for which order is lost at the core of the topological defect. In the presence of an applied magnetic field above a critical value, as a consequence of the Meissner effect  in type-II charged Bose-Einstein condensates, magnetic field is expelled from the bulk Bose-Einstein condensed system and into these magnetic topological defects.

Our primary interest in discussing 4D QHE phenomena is to develop a framework within which to approach a description of topological-order in 4D Bose-Einstein condensates, that are characterized by quaternion order parameters. Just as thin-film complex Bose-Einstein condensates must become topologically-ordered to achieve a phase-coherent ground state, as opposed to the usual Landau mechanism of spontaneous symmetry breaking~\cite{diamantini_topological_2008}, 4D quaternion Bose-Einstein condensates must follow a path of topological-ordering towards the orientationally-ordered ground state. {In this article, we suggest that the phase diagram of 4D quaternion Bose-Einstein condensates may be reproduced by making use of non-Abelian Yang-Mills topological terms~\cite{charap_gravitational_1977, atiyah_topological_1978, rivier_gauge_1990, rivier_line_1982}. This is a higher-dimensional analogue to 2D complex JJAs, for which the phase diagram (superconductor/superinsulator quantum phase transition) may be reproduced by mapping the planar JJA onto an Abelian gauge theory with a mixed Chern-Simons topological term~\cite{diamantini_gauge_1996, diamantini_topological_2008} that describes charge-vortex coupling.}

The remainder of this article is organized as follows. 2D quantum Hall effects, that can occur in systems of particles with $U(1)$ symmetry, are reviewed in Section~\ref{sec:2dqhe}. Firstly, in Section~\ref{sec:2deg}, QHE in 2D electron gases are discussed in order to introduce the relevance of $1^\text{st}$ Chern class topological invariants in 2D topologically-ordered phases. This understanding is then applied to consider an interpretation of dual frustrated superconductor and superinsulator ground states of 2D Josephson junction arrays. 4D QHE, and their relationship to dual frustrated ground states of quaternion Bose-Einstein condensates in 4D, are then introduced in Section~\ref{sec:4dqhe}.

\section{Theory of 2D Quantized Hall Effect and Superconductor-to-superinsulator transition in 2D Josephson junction arrays}
\label{sec:2dqhe}

\subsection{2D QHE in electron gases}
\label{sec:2deg}

Hall effects (HE) in conductors of electric charge occur in the presence of a transverse applied magnetic field $\textbf{B}$, such that a {Lorentz force} is exerted on a current $I$ flowing longitudinally in the conductor. In the HE, the Lorentz force separates charge carriers by their sign $(+,-)$, and this builds up a transverse ``Hall voltage'' between the conductor's lateral edges. Hall conductance is the longitudinal current divided by this transverse Hall voltage, and is not quantized in ``bulk'' 3D electron gases.

In contrast to Hall effects that occur in ``bulk'' 3D electron gases, Hall conductance in two-dimensional electron gases (2DEG) is quantized at low-temperatures by the strength of the imposed perpendicular magnetic field~\cite{laughlin_quantized_1981}. Precise quantization of Hall conductance in 2DEG, i.e., 2D quantum Hall effects (QHEs), holds regardless of the specifics of the experimental materials. This implies that QHEs have their origins in {topological ordering}~\cite{avron_topological_2003}. 

\begin{figure}[t!]
  \centering
\includegraphics[scale=.2]{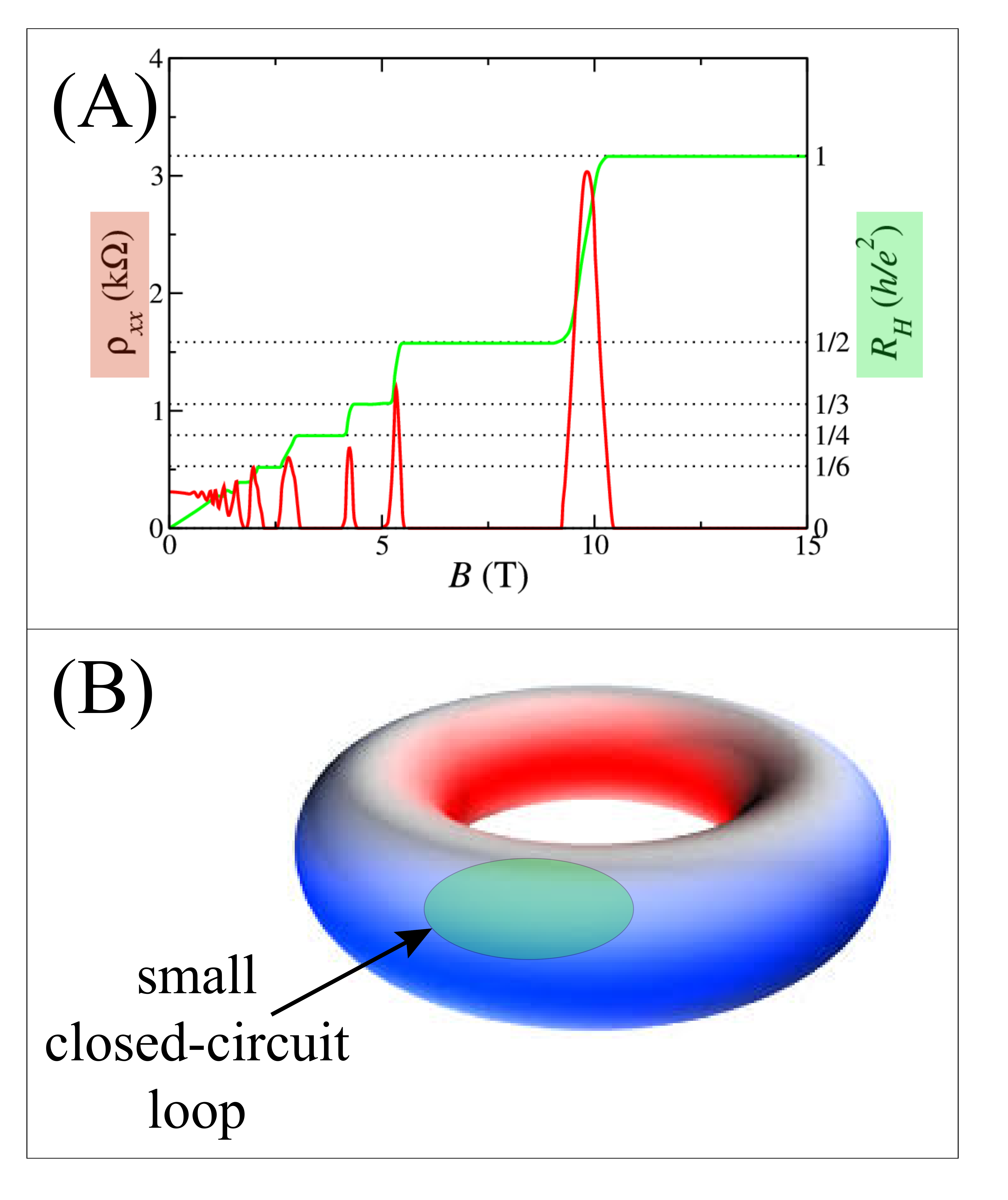}
\caption{(A) At low-enough temperatures, in 2DEG, Hall resistance is quantized as a function of a transverse magnetic field $B$; Hall resistance plateaus are precisely equal to  $R_H=I_x/V_y=h/n e^2$, where $n$ is the number of transferred electrons (Chern number). [Adapted from Ref.~\onlinecite{cooper_experimental_2012}]. (B) Curvature over an entire 2D toroidal surface (single genus) vanishes. However, the local curvature is positive in sphere-like regions (blue), negative near the hole (red) and the top/bottom circles have zero local curvature (grey). The $1^\text{st}$ Chern class topological invariants are determined by considering the angular mismatch of parallel transport, on making a small closed-circuit loop on the $T^2$ surface that encloses an area $dA$. }
\label{fig:qhall}
\end{figure}

In particular, Hall conductance in 2DEGs is quantized in integer values of the quantum of electrical Hall conductance: 
\begin{equation}
\sigma_H = n\cdot e^2/h, 
\label{eqn:hall_conductance}
\end{equation}
where $e$ and $h$ are the electronic charge and Planck's constant, and $n$ is the number of electrons transferred between the lateral edges of the conductor by a Hall current~\cite{laughlin_quantized_1981, avron_topological_2003}. Figure~\ref{fig:qhall} A shows a schematic of quantized Hall resistance (inverse of Hall conductance) in a 2DEG, at low-temperatures, as a function of applied magnetic field~\cite{avron_topological_2003, paalanen_quantized_1982}. It is well-known that quantization of plateaus in Hall resistance ($R_H$) may be interpreted using {Chern number} topological invariants~\cite{laughlin_quantized_1981, thouless_quantized_1982, bellissard_noncommutative_1994, avron_charge_1994} that belong to the $1^{st}$ Chern class.

Specifically,  $n$ is the average number of electrons transferred between the lateral edges of a conductor by a Hall current. This value is given by the Chern number topological invariant ($1^\text{st}$ Chern class), that characterizes the Hall-effect system~\cite{avron_topological_2003}. An elegant interpretation of the quantization of Hall conductance can be made by considering curvature of two-dimensional surfaces~\cite{laughlin_quantized_1981}. This is accomplished by considering the \emph{topology} of the Hall-effect Hamiltonian~\cite{avron_topological_2003}, which depends on two angular parameters $H(\Theta, \Phi)$ where: $\Theta$ is the $emf$ that drives the Hall current, and $\Phi$ is related to the ammeter that measures the Hall current~\cite{avron_topological_2003}. By gauge invariance~\cite{laughlin_quantized_1981}, the Hall-effect Hamiltonian is periodic in both of these parameters such that the Hamiltonian has the effective {topology} of a two-dimensional torus ($T^2$).

The Gauss-Bonnet-Chern formula (Ref.~\onlinecite{chern_curvatura_1945}) relates the geometry of Hall-effect eigenstates, parameterized by $\Theta$ and $\Phi$, to the torus topology of the Hall-effect Hamiltonian. This enables the definition of the Chern number topological invariant. The Gauss-Bonnet-Chern expression has the form:
\begin{equation}
\frac{1}{2\pi}\int_\mathcal{M} K dA = C_1.
\label{eqn:gauss_bonnet}
\end{equation}
The integral is over $\mathcal{M}=T^2$, a compact surface without a boundary, $K$ is the {adiabatic curvature} and $C_1$ is the Chern topological invariant that belongs to the $1^\text{st}$ Chern class. While the left side of Eqn.~\ref{eqn:gauss_bonnet} is geometric, and is therefore not implicitly quantized, the right side is quantized as the integer Chern number topological invariant $C_1$.

In order to determine the value of the $1^\text{st}$ Chern class invariant one must consider the closed-loop (i.e., the boundary of the green area in Fig.~\ref{fig:qhall} B) as the boundary of an ``inside'' area (enclosed by the loop) and as the boundary of an ``outside'' area (the rest of the surface). Failure of parallel transport is determined by considering the integrals of local curvature over the ``inside'' and ``outside'' areas (an angular mismatch, $K\,\, dA$). These two integrals agree up to an integer multiple of $2\pi$, and the Chern number invariant~\cite{avron_topological_2003, xiao_berry_2010} is the difference between them divided by $2\pi$. As the area enclosed by the loop is shrunk to zero, the Chern number invariant becomes the vanishing Euler characteristic of a torus.

Chern numbers are always integer values, because small changes in the Hall-effect Hamiltonian result in only small changes of {adiabatic curvature} and do not change the Chern number that characterizes the eigenstate. Only large changes in applied magnetic field (large deformations) of the Hamiltonian cause the system to cross over to other eigenstates~\cite{avron_topological_2003}, changing the Chern number topological invariant. Large deformations that allow the system to change eigenstates cause the system to undergo ``level-crossings'' that enable transitions between Chern number plateaus (see Fig.~\ref{fig:qhall} A). Chern number topological invariants, that identify a particular eigenstate of the Hall-effect Hamiltonian, actually quantize the average charge transported in the Hall effect in 2DEGs, i.e,. $C_1$ is $n$ (Eqn.~\ref{eqn:hall_conductance}).

\subsection{Superconductor/superinsulator quantum phase transition}
\label{sec:2D_JJA}

Like 2D quantum Hall effect systems, topological-order is a feature of complex Bose-Einstein condensates~\cite{diamantini_topological_2008} that exist in ``restricted dimensions'' (2D/1D) and that develop as Josephson junction arrays (JJAs) at finite temperatures. It is well-known that, in these scenarios, a phase-coherent superconductor ground state is obtained by a defect-driven Berezinskii-Kosterlitz-Thouless transition rather than by conventional spontaneous symmetry breaking. Relevant topological point defects in these scenarios are magnetic vortices, that belong to the fundamental homotopy group $\pi_1(S^1)$. By electric-magnetic duality, these topological charges are dual to electrical charges (Cooper pairs) and an Abelian topological Chern-Simons term~\cite{diamantini_gauge_1996, diamantini_topological_2008} governs charge-vortex coupling. Ultimately, this electric-magnetic duality leads to a quantum phase transition between dual phase-coherent superconductor and phase-incoherent superinsulator ground states~\cite{diamantini_gauge_1996, diamantini_topological_2008} at a self-dual critical point.

At the self-dual point, i.e., at the superconductor-to-superinsulator quantum phase transition, the dual Berezinskii-Kosterlitz-Thouless transition lines that allow for the existence of the dual phase-coherent and phase-incoherent ground states become entirely suppressed to 0 K. The topological-order in these dual ground states of JJAs, that may be realized in the vicinity of the superconductor-to-superinsulator quantum phase transition, have previously been discussed in terms of quantum Hall phases~\cite{diamantini_gauge_1996, diamantini_topological_2008} in the presence of either magnetic of charge frustration. Topological-order of dual frustrated ground states of JJAs, described in terms of dual Chern class invariants ($1^{st}$ Chern class), are reviewed in Sections~\ref{sec:2d_toposuperconductor} and ~\ref{sec:2d_toposuperinsulator}. This topological framework for dual topologically-ordered ground states of 2D JJAs  is then extended to four-dimensional quaternion Bose-Einstein condensates in Section~\ref{sec:4dqhe}.

   \begin{figure*}
  \centering
\includegraphics[width=\textwidth]{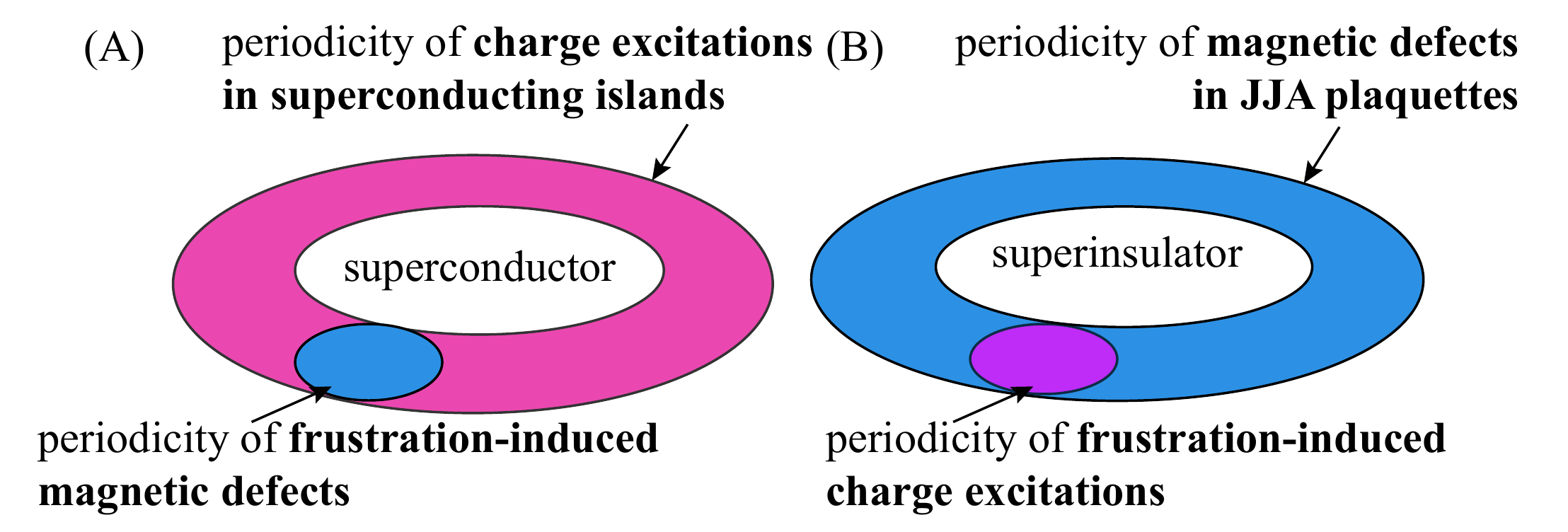}
\caption{(A) Topologically-ordered superconductors (phase-coherent) ground states of 2D/1D JJAs  may be characterized using $1^\text{st}$ Chern class invariants~\cite{diamantini_gauge_1996}. Torus topology ($T^2$) is generated by the periodicity of superconducting islands~\cite{diamantini_topological_2008}. Unfrustrated topologically-ordered superconducting ground states identify with a trivial Chern number topological invariant, and non-zero Chern numbers identify the degree of magnetic frustration in the vicinity of the superconductor-to-superinsulator transition. (B) By duality, topological-order in 2D superinsulating (phase-incoherent) ground states of JJAs may be understood within the $1^{st}$ Chern class by imposing torus topology that is generated by the periodicity of signed magnetic vortices (in plaquettes). In this dual limit, Chern number topological invariants characterize the degree of charge frustration~\cite{diamantini_topological_2008}.}
\label{fig:dual_tori}
\end{figure*}

     \begin{figure}[b!]
  \centering
\includegraphics[scale=1.1]{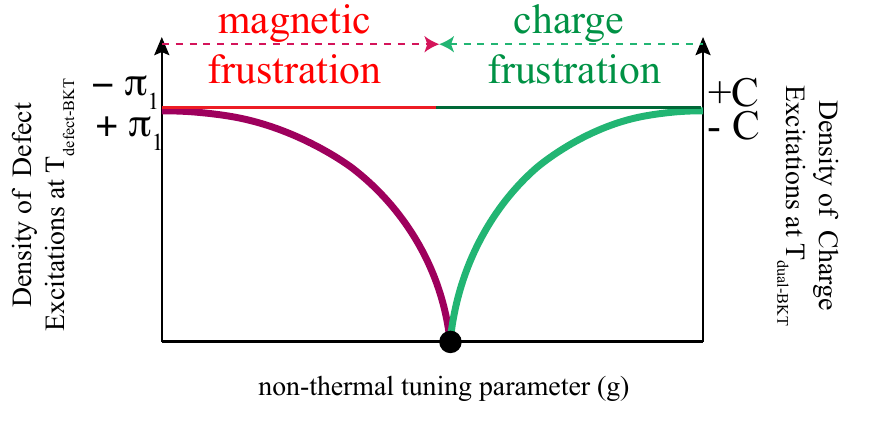}
\caption{{Density of magnetic topological point defects and charge excitations, as a function of the non-thermal tuning parameter $g$, in the vicinity of the ``self-dual'' critical point ($g=g_C$) at the relevant Berezinskii-Kosterlitz-Thouless transition temperature ($T=T_\text{defect-BKT}$ or $T=T_\text{dual-BKT}$).}}
\label{fig:density_excitations}
\end{figure}

\subsubsection{2D topologically-ordered superconductor}
\label{sec:2d_toposuperconductor}

In JJAs, low-temperature phase-coherent superconducting states are obtained by a defect-driven BKT transition of magnetic vortex point defects ($\pi_1(S^1)$) that interact logarithmically. In the absence of an applied magnetic field, concentrations of magnetic vortices with opposite signs are equal and all topological defects form bound pairs at $T_\text{BKT}<T_\text{BEC}$. The ground state is a perfectly topologically-ordered phase-coherent superconductor, that is free of topological defects because all low-energy bound pairs come together and annihilate on approaching 0 K. By considering that the JJA is fabricated with superconducting islands situated on a square planar lattice~\cite{diamantini_topological_2008}, and by imposing doubly periodic boundary conditions on the lattice, the system acquires torus topology. The ground state of the un-frustrated topologically-ordered superconductor, that forms by a perfect defect-driven BKT transition, can be characterized trivial Chern number ($1^{st}$ Chern class).

The presence of a transverse magnetic field biases the concentrations of topological magnetic point defects towards those of a single sign~\cite{teitel_josephson-junction_1983, gantmakher_superconductor-insulator_2010}. This suppresses the BKT transition, and the superconducting ground state now consist of a periodic configuration of magnetic vortices in the scalar phase angles across the phase-coherent JJA. As a result, by tuning an external parameter (magnetic field strength) superconductivity can be destroyed entirely as frustration-induced magnetic vortices that persist to the ground state begin to overlap (at the self-dual critical point). This superconductor-to-superinsulator quantum phase transition, at critical magnetic frustration, has been observed experimentally~\cite{eckern_quantum_1989, fazio_charge-vortex_1992, diamantini_gauge_1996, poran_quantum_2017} in 2D JJAs near 0 K. 

The degree of magnetic frustration is measured via the periodicity of the frustration-induced magnetic defects that persist to the ground state of the JJA~\cite{teitel_josephson-junction_1983}, in the presence of a transverse applied magnetic field. Specifically, in the presence of a transverse magnetic field, signed magnetic point defects form a periodic ground state with the unit cell $\textbf{q}\times \textbf{q}$ where $\textbf{q}$ is a count of the number of JJA plaquettes. Because of the periodicity of frustration-induced topological defects in the ground state, magnetically frustrated 2D superconductors can be viewed as topologically-ordered quantum Hall states~\cite{diamantini_gauge_1996} (that expels magnetic flux), classified by a non-zero Chern class topological invariant ($1^\text{st}$ Chern class). Figure~\ref{fig:dual_tori} A depicts a magnetically frustrated 2D superconducting ground state, with torus topology~\cite{diamantini_topological_2008}, that has a non-trivial $1^\text{st}$ Chern class topological invariant that applies globally.

\subsubsection{2D topologically-ordered superinsulator}
\label{sec:2d_toposuperinsulator}

As a consequence of electric-magnetic duality, JJAs that form in the range of dominant kinetic energy (of $O(2)$ quantum rotor models) can be described in terms of a lattice of signed magnetic vortices~\cite{diamantini_gauge_1996} (within each plaquette) -- instead of electrically charged superconducting islands.  In this dual limit that can be realized in two-dimensions, logarithmic interactions are realized between electrical charges within a real 2D Coulomb gas (of Cooper pairs). It follows that a dual Berezinskii-Kosterlitz-Thouless transition should occur at a critical finite temperature~\cite{mooij_unbinding_1990}. In the absence of charge frustration, a phase-incoherent ordered ground state forms at low-enough temperatures for which all condensed charged particles (Cooper pairs) become thermally pinned as charge-less dipoles within separate islands~\cite{mooij_unbinding_1990, baturina_superinsulatorsuperconductor_2013}. Such a ground state is free of frustration-induced charge excitations and (hence) has a trivial ``dual'' Chern invariant ($1^\text{st}$ Chern class).

Uniform charge frustration induces a background of elementary charge excitations that are free of dipole confinement~\cite{fazio_charge-vortex_1992}. These frustration-induced charge excitations form a {periodic} arrangement that is {dual} to the frustration-induced vortices in superconducting ground states in the presence of a transverse magnetic field~\cite{fazio_charges_2013}. Figure~\ref{fig:dual_tori} B depicts a superinsulating ground state of a JJA, for which the periodicity of magnetic point defects in plaquettes generates torus topology. The degree of uniform charge frustration is measured as a non-trivial Chern class invariant ($1^{st}$ Chern class). The case of charge-frustration shown in Figure~\ref{fig:dual_tori} B is dual to the magnetically-frustrated case (Figure~\ref{fig:dual_tori} A).

{At a critical value of charge frustration, i.e., at the quantum ({self-dual}) critical point, all condensed particles are liberated from charge-less dipoles and the low-temperature state is no longer one of superinsulation. Likewise, in the range of dominant potential energy of the $O(2)$ quantum rotor model, increasing magnetic frustration towards a critical value (by duality, Section~\ref{sec:2d_toposuperconductor}) skews the concentrations of magnetic point defects entirely towards those of a single sign in order to accommodate the applied transverse magnetic field~\cite{teitel_josephson-junction_1983, gantmakher_superconductor-insulator_2010}. Figure~\ref{fig:density_excitations} depicts a schematic of the biased nature of magnetic defect and charge excitation plasmas, just below $T_\text{BEC}$, in the presence of magnetic or charge frustration. In these scenarios of magnetic and charge frustration, in the vicinity of the ``self-dual'' critical point, excess signed charge excitations (topological magnetic or electronic) persist to the ground state as a periodic arrangement~\cite{fazio_charges_2013}. At the ``self-dual'' critical point, the plasmas of magnetic point defects and Cooper pairs that form at temperatures just below $T_\text{BEC}$ are entirely skewed towards excitations of a single sign. At this point, no bound pairs of charge excitations can form such that the ground state is neither one of superconductivity nor one of superinsulation.   }

     \begin{figure}[b!]
  \centering
\includegraphics[scale=.5]{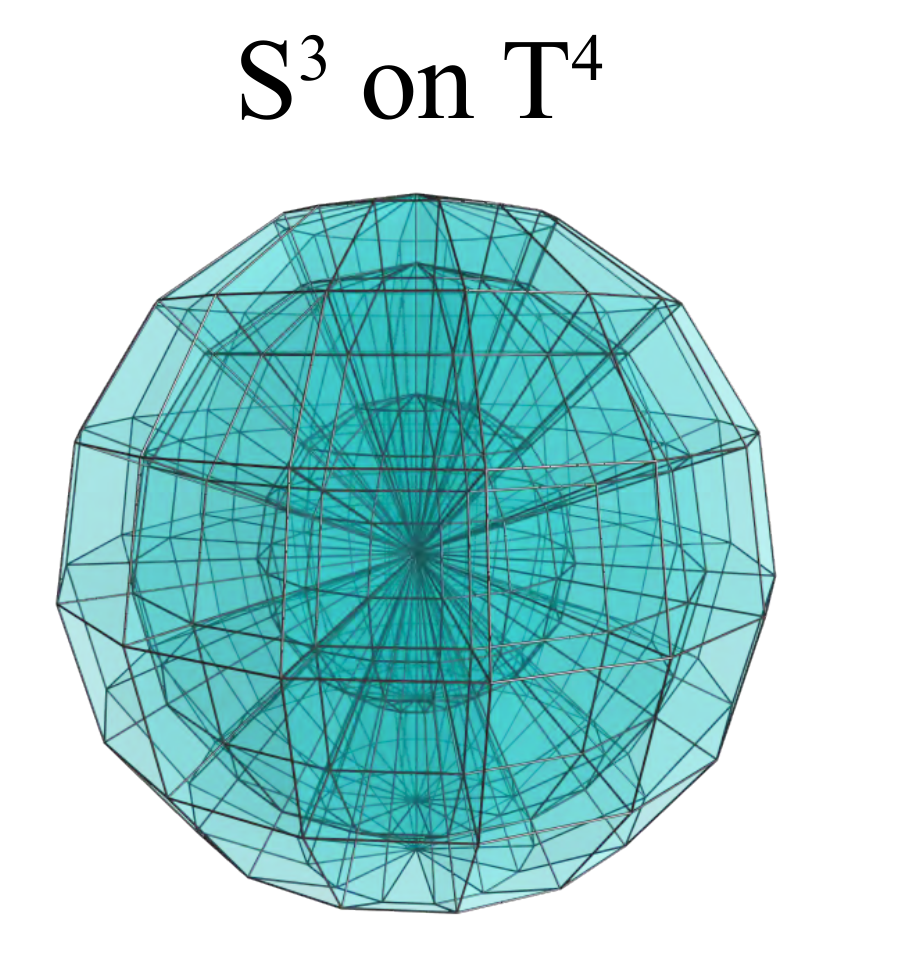}
\caption{When a closed-hypersphere is defined on the surface of a four-dimensional torus ($T^4$), the boundary can be taken as either an ``inner'' or ''outer'' hyper-volume; the difference between the two hyper-volumes (``inner'' and ``outer''), divided by $4\pi^2$, is the value of the  $2^\text{nd}$ Chern class topological invariant.}
\label{fig:paralleltransport}
\end{figure}

   \begin{figure*}
  \centering
\includegraphics[width=\textwidth]{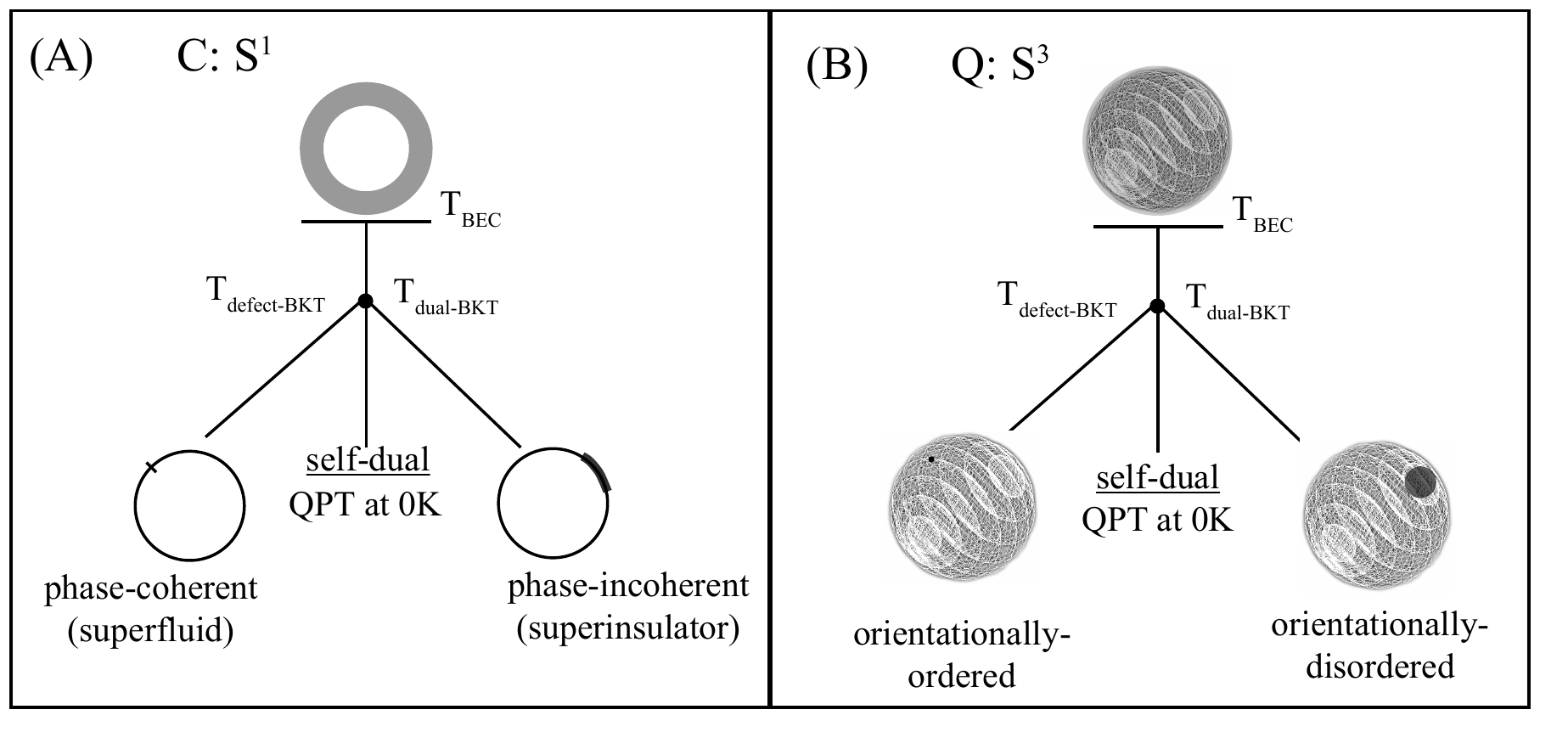}
\caption{(A) In ``restricted dimensions'' 2D/1D, a finite amount of undercooling is required below the bulk Bose-Einstein condensation critical temperature ($T_\text{BEC}$) as a consequence of the prevention of spontaneous symmetry breaking (Mermin-Wagner). In these systems, a duality is realized between magnetic vortices (misorientational fluctuations, topological charges) and Cooper pairs (electrical charges) that leads to the manifestation of dual phase-coherent (superconductor) and phase-incoherent (superinsulator) ground states. These dual ground states are separated by a {quantum critical point}, that is the end-point of the dual BKT transition lines. (B) Quaternion ordered systems that exist in 4D/3D are considered to exist in ``restricted dimensions.'' These ordered systems must undercool below $T_\text{BEC}$. Orientationally-ordered and orientationally-disordered ground states are anticipated on either side of a quantum phase transition, at a ``self-dual'' quantum critical point, as a consequence of a duality that is exhibited between third homotopy group topological point defects and condensed particle charges. {In application to solidification~\cite{gorham_topological_2019}, this ``self-dual'' critical point has previously been identified as the Kauzmann point~\cite{kauzmann_nature_1948, stillinger_kauzmann_2001} that occurs at finite temperatures and marks an ``ideal glass transition.''} }
\label{fig:tree_S1}
\end{figure*}

\section{4D QHE and Quantum Critical Point (``self-dual'') in 4D Bose-Einstein condensates}
\label{sec:4dqhe}

A generalization of 2D quantum Hall effects, that can occur in 2D systems of electron particles with $U(1)$ symmetry (Section~\ref{sec:2dqhe}), has been made~\cite{zhang_four-dimensional_2001, bernevig_eight-dimensional_2003, lohse_exploring_2018} to four-dimensional systems where particles have $SU(2)$ symmetry. The $2^\text{nd}$ Chern class topological invariants are relevant in these scenarios (see Fig.~\ref{fig:chern}). A generalized Gauss-Bonnet-Chern formula~\cite{chern_simple_1944, chern_curvatura_1945}, that applies in these scenarios, extends Eqn.~\ref{eqn:gauss_bonnet} to compact four-dimensional surfaces ($\mathcal{M}$): 
\begin{equation}
\frac{1}{(2\pi)^2}\int_\mathcal{M} Pf(\mathcal{M}) dA = C_2,
\label{eqn:gauss_bonnet_gen}
\end{equation}
where $Pf(\mathcal{M})$ is the Pfaffian curvature of $\mathcal{M}$, and $C_2$ is the Chern class topological invariant ($2^{nd}$ Chern class). The $2^\text{nd}$ Chern class topological invariants are identified by considering the failure of parallel transport on four-dimensional surfaces, for which a hyperspherical boundary on $T^4$ (Fig.~\ref{fig:paralleltransport}) takes the place of the small closed-circuit loop on $T^2$ (Fig.~\ref{fig:qhall} B). Just like in the case of $1^\text{st}$ Chern class topological invariants, this hyperspherical boundary can be taken as either: enclosing an ``inside'' hypervolume or, defining an ``outside'' hypervolume that accounts for the rest of the surface $\mathcal{M}$. The difference in the two integrals of local curvature (Eqn.~\ref{eqn:gauss_bonnet_gen}, ``inside'' and ``outside'') must agree up to an integer multiple of $4\pi^2$, and the difference between them divided by $4\pi^2$ is the $2^\text{nd}$ Chern class invariant. 

Just like 4D QHE phenomena, in systems of particles with $SU(2)$ symmetry, Bose-Einstein condensates that are characterized by a quaternion order parameter and that exist in four-dimensions are topologically-ordered at low-temperatures. This is a consequence of the fact that quaternion numbers form a four-dimensional compact Lie group, for which a given quaternion number is characterized by three separate scalar phase angle parameters. Thus, quaternion Bose-Einstein condensates allow for the existence of third homotopy group topological defects that play the role of magnetic vortices in superconductors. Third homotopy group topological defects behave as points in four-dimensions and, such quaternion ordered systems that exist in four- or three- dimensions are unable to undergo a conventional disorder-order transition at finite temperatures. Just like ordered systems that exhibit a continuous symmetry and that exist in 2D and 1D (Mermin-Wagner), these systems will follow an alternative path of topological ordering towards the orientationally-ordered ground state that is defect-driven.

Four-dimensional quaternion ordered systems are higher-dimensional generalization of thin-film complex Josephson junction arrays~\cite{gorham_topological_2019, gorham_crystallization_2018}. Just as dual phase-coherent and phase-incoherent ground states of 2D JJAs may be obtained, by dual Berezinskii-Kosterlitz-Thouless transitions (Figure~\ref{fig:tree_S1} A), a transition between orientationally-ordered and orientationally-disordered ground states of 4D quaternion ordered systems is anticipated at a self-dual critical point (Figure~\ref{fig:tree_S1} B). This is a consequence of a duality that is realized between condensed particles and third homotopy group topological point defects. The dual frustrated ground states, that can develop at low-enough temperatures on either side of a quantum ({self-dual}) critical point, are discussed in Sections~\ref{sec:charged_4d} and ~\ref{sec:topological_4d} in terms of 4D quantum Hall phases.

\subsection{4D topologically-ordered ``superconductor''}
\label{sec:charged_4d}

In four-dimensional quaternion Bose-Einstein condensates, orientationally-ordered ground states may develop at finite temperatures via a defect-driven Berezinskii-Kosterlitz-Thouless transition of third homotopy group topological point defects. In the absence of ``magnetic'' frustration, the plasma of third homotopy group topological defects that develops is perfectly balanced such that a perfect defect-driven BKT transition is anticipated at $T_\text{defect-BKT}<T_\text{BEC}$. Such an orientationally-ordered ground state is defect-free (unfrustrated), as all low-energy bound topological defect pairs will come together and annihilate as the temperature is lowered towards 0 K. In order to consider this low-temperature ordered system as a 4D quantum Hall phase, it is necessary to impose periodic boundary conditions on the four-dimensional lattice of quaternion ordered ``islands.'' In this way, the system acquires the topology of a four-dimensional torus ($T^4$). Perfectly orientationally-ordered ground states, i.e., in the absence of ``magnetic'' frustration, are then identified by a vanishing Chern invariant ($2^\text{nd}$ Chern class).

Uniform ``magnetic'' frustration biases the plasma of third homotopy group topological defects (misorientational fluctuations) towards those of a single sign, which suppresses the defect-driven BKT transition temperature. Ultimately, with a critical amount of frustration, the topological-ordering transition towards an orientationally-ordered ground state is entirely suppressed to 0 K at a ``self-dual'' critical point. Excess signed third homotopy group topological point defects are unable to form bound pairs, and are forced into the orientationally-ordered ground state as a periodic arrangement. This is a higher-dimensional analogue to the frustration-induced magnetic vortices that persist to phase-coherent ground states of 2D JJAs, in the presence of an applied transverse magnetic field~\cite{teitel_josephson-junction_1983} (Abrikosov flux lattice).

While in the absence of ``magnetic'' frustration the ground state has a vanishing Chern number invariant, frustrated topologically-ordered ground states are characterized by non-trivial Chern number invariants. This can be seen directly by mapping the periodicity of frustration-induced third homotopy group topological point defects onto the representation of the orientationally-ordered ground state as a four-dimensional torus $T^4$. This is analogous to the magnetically-frustrated ground state of a 2D JJA, shown in Figure~\ref{fig:dual_tori} A. The limit of critical ``magnetic'' frustration corresponds to the ``self-dual'' critical point at which the concentrations of third homotopy group point defects are entirely biassed towards those of a single sign, and signed defect cores begin to overlap in the ground state. At this self-dual critical point, the orientationally-ordered low-temperature ground state is entirely suppressed. With critical frustration, the system may be identified with the largest possible $2^\text{nd}$ Chern class invariant.

\subsection{4D topologically-ordered ``superinsulator''}
\label{sec:topological_4d}

As a consequence of a duality between third homotopy group point defects and condensed particles that is realized for quaternion ordered systems in four-dimensions, in the limit of small coupling between quaternion ordered islands, an orientationally-disordered ground state can be obtained. This ground states is described in terms of a four-dimensional lattice of signed third homotopy group defects -- instead of as a lattice of charged particles on quaternion ordered islands. Such an orientationally-disordered ground state may be realized via a dual Berezinskii-Kosterlitz-Thouless transition within a gas of Cooper pair-like charges, in analogue to the formation of a ``superinsulating'' ground state of JJAs in 2D/1D.

In the absence of interaction energy between neighboring ordered ``islands,'' a maximally orientationally-disordered ground state is achieved that is not frustrated in that all Cooper pairs form charge-less dipoles at the dual-BKT transition temperature. Such an unfrustrated orientationally-disordered ground state is characterized by a trivial ``dual'' Chern number topological invariant ($2^\text{nd}$ Chern class). Incorporation of uniform interaction energy throughout the system leads to uniform frustration of orientationally-disordered ground states, by allowing for a periodic arrangement of aligned charge excitations to persist to 0 K. Uniform frustration in orientationally-disordered ground states is measured by a non-trivial ``dual'' Chern number belonging to the $2^\text{nd}$ Chern class.

\section{Summary and Discussion}

In this article, we have applied the generalization of quantum Hall effects from 2D to 4D in order to consider topological-ordering phenomena in quaternion Bose-Einstein condensates that exist in ``restricted dimensions'' (i.e., 4D/3D).  In charged Bose-Einstein condensates, relevant applied magnetic fields are pushed into first (complex) and third (quaternion) homotopy group topological defects -- for which the defect core is in the ``normal'' state. For charged Bose-Einstein condensates that exist in ``restricted dimensions,''  these magnetic defects are spontaneously generated at finite temperatures and prevent spontaneous symmetry breaking (Mermin-Wagner). 

In these scenarios of Bose-Einstein condensation in ``restricted dimensions,'' electric-magnetic duality leads to a quantum phase transition at a ``self-dual'' critical point between phase-coherent (superconductor) and phase-incoherent (superinsulator) ground states. These dual ground states are obtained by dual Berezinskii-Kosterlitz-Thouless topological ordering transitions (of magnetic point defects and of Cooper pair charges). The dual BKT transition lines meet at the ``self-dual'' critical point, which marks the quantum phase transition between dual ground states.

{Four-dimensional quaternion Bose-Einstein condensates are a direct higher-dimensional analogue to thin-film complex Josephson junction arrays. The Hamiltonians that describe these complex and quaternion ordered systems, that exist in ``restricted-dimensions,'' map directly to 2D (complex) and 4D (quaternion) torus topology by the imposition of doubly and quadruply periodic boundary conditions. This enables the identification of frustrated dual ground states, on either side of the ``self-dual'' critical point, in terms of global topological invariants that belong to the $1^\text{st}$ Chern class (complex) or $2^\text{st}$ Chern class (quaternion).}

Notably, the authors have previously applied this framework for topological ordering -- based on a quaternion orientational order parameter -- to consider the formation of crystalline and glassy solid states from undercooled fluids (Ref.~\onlinecite{gorham_topological_2019}). However, in these 3D solid state systems (uncharged), the focus has mainly surrounded the topological-ordering of disclination line defects (rotational) into dislocations (translational) -- instead of the topological-ordering of third homotopy group defects. This common approach has proven attractive because line defects are apparent in the solid state, while third homotopy group defects are not. In the solid state, {geometrical frustration} (instead of {magnetic frustration}) has been utilized as the non-thermal tuning parameter that drives the system towards a ``self-dual'' critical point. {This critical point has been previously introduced by the authors, in application to the solid state, as the Kauzmann point that occurs at finite temperatures and marks an ``ideal glass transition'' (crystalline-to-glass transition).}

\section{Acknowledgment}

D.E.L. and C.S.G. acknowledge support from the ALCOA Chair in Physical Metallurgy.

\bibliography{\jobname}

\begin{thebibliography}{10}

\bibitem{lohse_exploring_2018}
M.~Lohse, C.~Schweizer, H.~M. Price, O.~Zilberberg, and I.~Bloch, ``Exploring
  4d {Quantum} {Hall} {Physics} with a 2d {Topological} {Charge} {Pump},'' {\em
  Nature}, vol.~553, pp.~55--58, Jan. 2018.
\newblock arXiv: 1705.08371.

\bibitem{diamantini_gauge_1996}
M.~C. Diamantini, P.~Sodano, and C.~A. Trugenberger, ``Gauge theories of
  {Josephson} junction arrays,'' {\em Nuclear Physics B}, vol.~474, no.~3,
  1996.

\bibitem{diamantini_topological_2008}
M.~C. Diamantini, P.~Sodano, and C.~A. Trugenberger, ``Topological order in
  frustrated {Josephson} junction arrays,'' {\em EPL (Europhysics Letters)},
  vol.~83, p.~21003, July 2008.

\bibitem{zhang_four-dimensional_2001}
S.-C. Zhang and J.~Hu, ``A four-dimensional generalization of the quantum
  {Hall} effect,'' {\em Science}, vol.~294, no.~5543, pp.~823--828, 2001.

\bibitem{bernevig_eight-dimensional_2003}
B.~A. Bernevig, J.~Hu, N.~Toumbas, and S.-C. Zhang, ``Eight-dimensional quantum
  {Hall} effect and “octonions”,'' {\em Physical review letters}, vol.~91,
  no.~23, p.~236803, 2003.

\bibitem{kraus_four-dimensional_2013}
Y.~E. Kraus, Z.~Ringel, and O.~Zilberberg, ``Four-dimensional quantum {Hall}
  effect in a two-dimensional quasicrystal,'' {\em Physical review letters},
  vol.~111, no.~22, p.~226401, 2013.

\bibitem{toulouse_principles_1976}
G.~Toulouse and M.~Kléman, ``Principles of a classification of defects in
  ordered media,'' {\em Journal de Physique Lettres}, vol.~37, no.~6, 1976.

\bibitem{gorham_topological_2019}
C.~S. Gorham and D.~E. Laughlin, ``Topological description of the
  solidification of undercooled fluids and the temperature dependence of the
  thermal conductivity of crystalline and glassy solids above approximately 50
  {K},'' {\em Journal of Physics: Condensed Matter}, vol.~31, p.~105701, Jan.
  2019.

\bibitem{charap_gravitational_1977}
J.~M. Charap and M.~J. Duff, ``Gravitational effects on {Yang}-{Mills}
  topology,'' {\em Physics Letters B}, vol.~69, pp.~445--447, Jan. 1977.

\bibitem{atiyah_topological_1978}
M.~F. Atiyah and J.~D.~S. Jones, ``Topological aspects of {Yang}-{Mills}
  theory,'' {\em Communications in Mathematical Physics}, vol.~61, no.~2, 1978.

\bibitem{rivier_gauge_1990}
N.~Rivier, ``Gauge theory and geometry of condensed matter,'' {\em Geometry in
  condensed matter physics}, 1990.

\bibitem{rivier_line_1982}
N.~Rivier and D.~Duffy, ``Line defects and tunnelling modes in glasses,'' {\em
  Journal de Physique}, vol.~43, no.~2, pp.~293--306, 1982.

\bibitem{laughlin_quantized_1981}
R.~B. Laughlin, ``Quantized {Hall} conductivity in two dimensions,'' {\em
  Physical Review B}, vol.~23, no.~10, p.~5632, 1981.

\bibitem{avron_topological_2003}
J.~E. Avron, D.~Osadchy, and R.~Seiler, ``A topological look at the quantum
  {Hall} effect,'' {\em Physics today}, vol.~56, no.~8, pp.~38--42, 2003.

\bibitem{paalanen_quantized_1982}
M.~A. Paalanen, D.~C. Tsui, and A.~C. Gossard, ``Quantized {Hall} effect at low
  temperatures,'' {\em Physical Review B}, vol.~25, no.~8, p.~5566, 1982.

\bibitem{thouless_quantized_1982}
D.~J. Thouless, M.~Kohmoto, M.~P. Nightingale, and M.~den Nijs, ``Quantized
  {Hall} conductance in a two-dimensional periodic potential,'' {\em Physical
  Review Letters}, vol.~49, no.~6, p.~405, 1982.

\bibitem{bellissard_noncommutative_1994}
J.~Bellissard, A.~van Elst, and H.~Schulz-Baldes, ``The noncommutative geometry
  of the quantum {Hall} effect,'' {\em Journal of Mathematical Physics},
  vol.~35, no.~10, pp.~5373--5451, 1994.

\bibitem{avron_charge_1994}
J.~E. Avron, R.~Seiler, and B.~Simon, ``Charge deficiency, charge transport and
  comparison of dimensions,'' {\em Communications in mathematical physics},
  vol.~159, no.~2, pp.~399--422, 1994.

\bibitem{cooper_experimental_2012}
D.~R. Cooper, B.~D’Anjou, N.~Ghattamaneni, B.~Harack, M.~Hilke, A.~Horth,
  N.~Majlis, M.~Massicotte, L.~Vandsburger, and E.~Whiteway, ``Experimental
  review of graphene,'' {\em ISRN Condensed Matter Physics}, vol.~2012, 2012.

\bibitem{chern_curvatura_1945}
S.-S. Chern, ``On the {Curvatura} {Integra} in a {Riemannian} {Manifold},''
  {\em Annals of Mathematics}, vol.~46, no.~4, 1945.

\bibitem{xiao_berry_2010}
D.~Xiao, M.-C. Chang, and Q.~Niu, ``Berry phase effects on electronic
  properties,'' {\em Reviews of modern physics}, vol.~82, no.~3, p.~1959, 2010.

\bibitem{teitel_josephson-junction_1983}
S.~Teitel and C.~Jayaprakash, ``Josephson-junction arrays in transverse
  magnetic fields,'' {\em Physical review letters}, vol.~51, no.~21, 1983.

\bibitem{gantmakher_superconductor-insulator_2010}
V.~F. Gantmakher and V.~T. Dolgopolov, ``Superconductor-insulator quantum phase
  transition,'' {\em Physics-Uspekhi}, vol.~53, no.~1, 2010.

\bibitem{eckern_quantum_1989}
U.~Eckern and A.~Schmid, ``Quantum vortex dynamics in granular superconducting
  films,'' {\em Physical Review B}, vol.~39, no.~10, p.~6441, 1989.

\bibitem{fazio_charge-vortex_1992}
R.~Fazio, A.~Van~Otterlo, G.~Schön, H.~S.~J. Van Der~Zant, and J.~E. Mooij,
  ``Charge-{Vortex} {Duality} in {Josephson} {Junction} {Arrays},'' {\em Helv.
  Phys. Acta}, vol.~65, no.~228, 1992.

\bibitem{poran_quantum_2017}
S.~Poran, T.~Nguyen-Duc, A.~Auerbach, N.~Dupuis, A.~Frydman, and O.~Bourgeois,
  ``Quantum criticality at the superconductor-insulator transition revealed by
  specific heat measurements,'' {\em Nature Communications}, vol.~8, 2017.

\bibitem{mooij_unbinding_1990}
J.~E. Mooij, B.~J. van Wees, L.~J. Geerligs, M.~Peters, R.~Fazio, and
  G.~Schön, ``Unbinding of charge-anticharge pairs in two-dimensional arrays
  of small tunnel junctions,'' {\em Physical Review Letters}, vol.~65, no.~5,
  1990.

\bibitem{baturina_superinsulatorsuperconductor_2013}
T.~I. Baturina and V.~M. Vinokur, ``Superinsulator–superconductor duality in
  two dimensions,'' {\em Annals of Physics}, vol.~331, 2013.

\bibitem{fazio_charges_2013}
R.~Fazio and G.~Schön, ``Charges and {Vortices} in {Josephson} {Junction}
  {Arrays},'' in {\em 40 {Years} of {Berezinskii}–{Kosterlitz}–{Thouless}
  {Theory}}, pp.~237--254, World Scientific, 2013.

\bibitem{kauzmann_nature_1948}
W.~Kauzmann, ``The {Nature} of the {Glassy} {State} and the {Behavior} of
  {Liquids} at {Low} {Temperatures}.,'' {\em Chemical Reviews}, vol.~43, no.~2,
  1948.

\bibitem{stillinger_kauzmann_2001}
F.~H. Stillinger, P.~G. Debenedetti, and T.~M. Truskett, ``The {Kauzmann}
  paradox revisited,'' {\em The Journal of Physical Chemistry B}, vol.~105,
  no.~47, pp.~11809--11816, 2001.

\bibitem{chern_simple_1944}
S.-S. Chern, ``A {Simple} {Intrinsic} {Proof} of the {Gauss}-{Bonnet} {Formula}
  for {Closed} {Riemannian} {Manifolds},'' {\em Annals of Mathematics},
  vol.~45, no.~4, pp.~747--752, 1944.

\bibitem{gorham_crystallization_2018}
C.~S. Gorham and D.~E. Laughlin, ``Crystallization in {Three}-{Dimensions}:
  {Defect}-{Driven} {Topological} {Ordering} and the {Role} of {Geometrical}
  {Frustration},'' {\em arXiv:1812.11265 [cond-mat]}, Dec. 2018.
\newblock arXiv: 1812.11265.

\end{thebibliography}

\end{document}